\documentclass[]{aa}
\usepackage{graphicx,txfonts,natbib}
\bibpunct{(}{)}{;}{a}{}{,}
\newcommand{\lya}{Ly$\alpha$}
\newcommand{\ha}  {H$\alpha$}
\newcommand{\ecs}{\hbox{erg cm$^{-2}$ s$^{-1}$}}
\begin{document}
   \title{A Lyman\,$\alpha$ emitter at $z = 6.5$ found with slitless
   spectroscopy\thanks{Based on observations obtained at the
   European Southern Observatory using the ESO Very Large Telescope on
   Cerro Paranal (ESO programs 066.A-0122, 068.A-0182, 272.A-5007).}}

   \author{J.D. Kurk\inst{1} \and A. Cimatti\inst{1} \and 
           S. di Serego Alighieri\inst{1} \and J. Vernet\inst{1} \and
           E. Daddi\inst{2} \and A. Ferrara\inst{1} \and B. Ciardi\inst{3}
          }

%  \offprints{kurk@arcetri.astro.it}

   \institute{INAF, Osservatorio Astrofisico di Arcetri, 
              Largo E. Fermi 5, 50125, Firenze
         \and
             European Southern Observatory,
             Karl-Schwarzschild-Strasse 2, 85748, Garching bei
             M\"unchen, Germany
         \and
             Max-Planck-Institut f\"ur Astrophysik,
             Karl-Schwarzschild-Strasse 1, 85748, Garching bei
             M\"unchen, Germany
             }

   \date{Received; accepted}

   \abstract{ We report the discovery of a Lyman\,$\alpha$ emitting
   galaxy at $z = 6.518$.  The single line was found in the 43
   arcmin$^2$ VLT/FORS field by slitless spectroscopy limited to the
   atmospheric window at $\lambda \sim 9100$\,\AA\ (sampling \lya\ at
   $6.4 < z < 6.6$).  Its counterpart is undetected in a deep $I$ band
   image and the line has an asymmetric appearance in a deeper
   follow-up spectrum.  There are no plausible line identifications
   except for Ly$\alpha$ with a flux of 1.9$\times 10^{-17}$\,\ecs\
   and rest frame equivalent width $> 80$\,\AA.  The lower limit to
   the star formation rate density at $z = 6.5$ derived from our
   complete sample is 5$\times$10$^{-4}$ M$_\odot$ yr$^{-1}$
   Mpc$^{-3}$, consistent with measurements in the Subaru Deep Field
   and Hubble Ultra Deep Field but approximately ten times higher than
   in the Large Area Lyman Alpha survey. This \lya\ emitter is among
   the very small sample of highest redshift galaxies known.

   }

   \maketitle
%
%________________________________________________________________

\section{Introduction}

While reaching for the stars, our arms are still growing longer.
Until a few years ago, the most distant objects known were QSOs at
$z\sim 5.8$ \citep{2000AJ....120.1167F} and a handful of Ly$\alpha$
emitters (LAEs) at $5 < z < 6$ \citep[][ for a review]
{1999PASP..111.1475S}.  However, advances in the past years with
respect to the red sensitivity of CCDs have enabled the detection of
galaxies at even higher redshifts \citep[][ for a
review]{2003JKAS...36..123T}.

The detection of $z > 6$ galaxies allows to study the star formation
rate in the early Universe, the modes of early galaxy formation and
the interplay between the first galaxies and the intergalactic medium
(IGM).  With the large sample of Lyman break galaxies (LBGs) now
available at $z \sim 3$ and $z \sim 4$, the star formation rate
density (SFRD) at these redshifts is quite well determined
\citep[e.g.][]{1999ApJ...519....1S}.  At higher redshifts, there have
been tentative determinations of the SFRD recently, but these are
still based on poor statistics and some uncertainties in the modeling
\citep{2004Natur.428..625H, Bunker2004, Ricotti2004}.  At high
redshift, we may encounter stars formed out of primordial elements,
which differ strongly from those formed from enriched material, due to
differences in initial mass function (IMF) and stellar temperatures.
These so--called \emph{Population III} (PopIII) stars are responsible
for the production of the first metals in the Universe
\citep{2003ApJ...589...35S, 2003A&A...397..527S}.  The redshift range
$6 < z < 7$ is a very intriguing time during cosmic evolution, when
hydrogen reionization is believed to be basically complete and the IGM
starts to be polluted with metals.  Even though it is now believed
that reionization of most of the hydrogen takes place earlier
\citep{2003ApJS..148..161K}, large scale structure may inhibit a
homogeneous distribution of ionizing radiation, introducing a large
cosmic variance in the reionization \citep{2000ApJ...542..535G,
2003MNRAS.343.1101C}.

As the \lya\ line can be a prominent feature of young star forming
galaxies, this line has been used extensively to search for distant
galaxies.  Blind spectroscopic searches are carried out from Earth
\citep[e.g.\ ][]{2001ApJ...560L.119E, 2004ApJ...603..414M} and from space
\citep{Rhoads2004a, Pirzkal2004}, but most surveys are performed by
narrow band imaging with ground based telescopes.  Using narrow band
filters sensitive to wavelengths in the atmospheric window at $\lambda
\sim 0.9\,\mu$m, corresponding to Ly$\alpha$ emission at $z \sim 6.5$,
and subsequent spectroscopy, three groups have detected galaxies at $z
= 6.5$. The first of these was found due to the lensing amplification
(4.5$\times$) by the cluster Abell 370 \citep{2002ApJ...568L..75H},
two were found by surveying the huge field of view (814 arcmin$^2$) of
the Subaru Suprime-Cam instrument \citep{2003PASJ...55L..17K} and one
during the course of the Large Area Lyman Alpha survey \citep[LALA,
][]{Rhoads2004}.  Detections of even higher redshift, very strongly
lensed, galaxies are reported by \citet[][ $z = 7$]{Kneib2004} and
\citet[][ $z = 10$] {2004A&A...416L..35P}.

We have followed a different method to find \lya\ emitting galaxies at
$z \sim 6.5$ and positively identify one galaxy at $z = 6.518$.
Throughout this Letter, we adopt a flat Universe with $\Omega_M =
0.3$, $\Omega_\Lambda = 0.7$ and $H_0 = 65$ km\,s$^{-1}$\,Mpc$^{-1}$.

%__________________________________________________________________

\section{Observations}

Our approach to find LAEs at $z \sim 6.5$ is to use ground based
slitless spectroscopy in combination with a medium band filter and
imaging carried out with FORS2 at the VLT of a
7\arcmin$\times$7\arcmin\ field with very low Galactic extinction
($E_{B-V}$=0.002) centered at RA = 04:01:55.9, Dec = -37:43:34
(J2000).  The filter (ESO's z\_SPECIAL+43, $z_{sp}$ from now on) has a
central wavelength of 9135\,\AA\ and FWHM of 193\,\AA\ and is
therefore sensitive to Ly$\alpha$ originating from $z = 6.433$ to $z =
6.592$.  The filter reduces the sky background and overlap problems by
providing \emph{short} spectra.  This technique provides a
completeness limit on line flux and the redshift of lines with enough
resolving power to separate possible contaminants from real LAEs.

% ESO's filter curve: FWHM 9052 - 9217 A, center at 9134 A, W 165 A

We used the holographic 600$z$ grism which delivers a resolution of
about 1400 for 1\arcsec\ objects.  Over a period of four months in the
winter of 2002/03, an exposure of 7.6 hours was collected.  Weather
conditions were excellent resulting in 0\farcs7 seeing.  Direct
imaging was carried out through the $z_{sp}$ filter, to determine the
sources of emission lines detected in the spectroscopy, and Bessel-$I$
filter, to determine the $I - z_{sp}$ colour.  A Gunn-$v$ image was
available from earlier observations of this field.  Total exposure
times were 6.33, 4.0 and 1.0 hours with seeing discs of 0\farcs7,
0\farcs7 and 0\farcs6 for the $z_{sp}$, $I$ and $v$ bands,
respectively.

Reduction was carried out in a standard way.  Flux calibration of the
spectroscopic image was provided by observations of two
spectrophotometric standards.  For the $z_{sp}$ filter, the literature
values were convolved with the sensitivity of the filter to obtain the
magnitude.  Three sigma limiting AB magnitudes for a 1\arcsec\
aperture are 27.0 and 27.5, respectively for the $z_{sp}$ and $I$
band.  Special calibration were carried out to determine the
distortion pattern of the FORS2 instrument using the movable slitlets
of FORS2 in direct and spectroscopic modes using seven slit
configurations to cover the entire field. These calibration images
allowed us to determine the region on the direct image where the
source of a spectrum or line on the spectroscopic image can be located
to within $\sim$1 pixel or 0\farcs25.

In november 2003, we have also carried out 2.1 hours of multi object
spectroscopy in Director's Discretionary Time (DDT) with FORS2 using
the 1028$z$ grism, which delivers a resolution of about 2600 for the
1\arcsec\ slits used in the mask.  The eight frames were offset by
3\arcsec\ in an ABBA sequence.  The observations were reduced in a
standard way, by subtraction of the overscan region and flat fielding
with screen flats.  Cosmic rays were rejected by combining the two
pairs of frames with the same offset.  Wavelength calibration and
rectification of the two--dimensional spectra were carried out by
measuring the position of about 10 sky lines.  The error in the
wavelength calibration is dominated by fringing which influences the
determination of the center of the sky lines and limits the wavelength
accuracy to 0.2\,\AA.
%Spectroscopy of LTT\,2415 \citep{1992PASP..104..533H} served to
%calibrate the flux.

\section{Results}

%
%                                                One column figure
%-----------------------------------------------------------
   \begin{figure} \centering
   \includegraphics[width=\columnwidth]{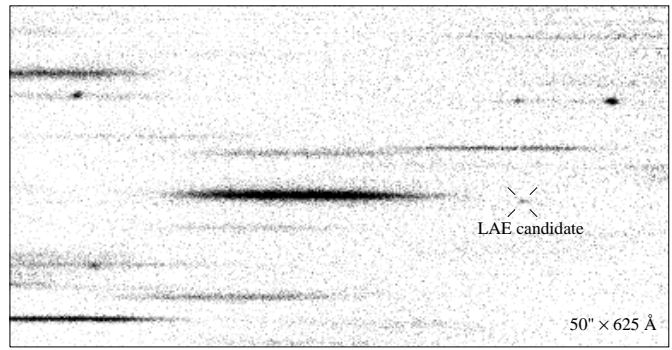} \caption{Part of
   the slitless image, including the targeted \lya\ candidate and an
   [\ion{O}{iii}] emitter at $z = 0.8$ at the top right.}
   \label{mos_image} \end{figure}
%
%______________________________________________________________

%
%                                                One column figure
%-----------------------------------------------------------
   \begin{figure} \centering
   \includegraphics[width=\columnwidth]{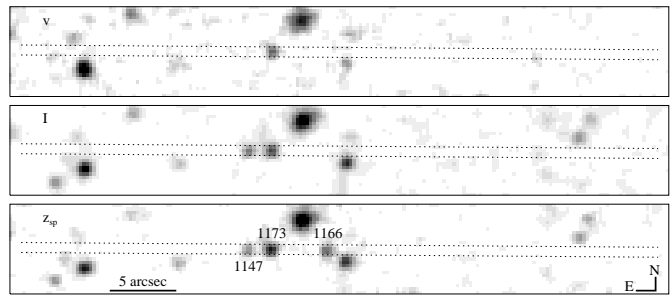} \caption{Images in
   the $v$, $I$ and $z_{sp}$ band of the region where possible
   counterparts of the emission line can be located (between the
   dotted lines).  The catalog numbers of the three possible
   counterparts are indicated.  Note that KCS 1166 is only detected
   in $z_{sp}$.}  \label{direct} \end{figure}
%
%______________________________________________________________

%
%                                                One column figure
%-----------------------------------------------------------
   \begin{figure} \centering
   \includegraphics[width=\columnwidth]{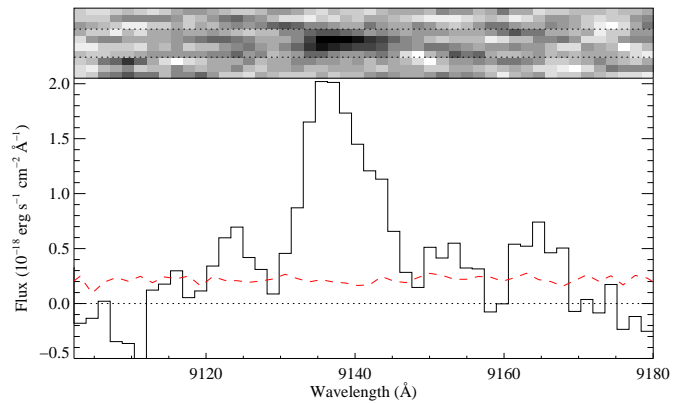}
   \caption{One and two dimensional slitless spectrum of the LAE
   candidate after smoothing by a 3 pixel boxcar along the dispersion
   direction to increase the s/n without reducing the resolution
   significantly.  The dotted lines indicate the extraction aperture
   (4 pixels or 1\farcs0).  The dashed line indicates the signal to
   noise per pixel (after smoothing).}  \label{slitless_lya}
   \end{figure}
%
%______________________________________________________________

In the slitless spectroscopic image, three authors (JK, AC and SSA)
have independently searched for emission lines by visual inspection.
Automatic extraction has been attempted but fails to detect emission
lines superposed on continua and results in many fake detections.  The
visual inspection has a completeness level which corresponds to line
fluxes of 1.0 to 2.0$\times$10$^{-17}$\,\ecs\ depending on the
(unknown) wavelength of the line (values stated are for a line at the
maximum of the filter transmission and half of it). Fig.\
\ref{mos_image} shows a small part of the slitless image, including a
LAE candidate which has a signal close to the completeness limit.  In
the resulting combined list, there are 161 lines, each of which can
originate from a counterpart within a region of about
50\arcsec$\times$1\arcsec\ on the direct image.  Taking all objects
detected by SExtractor in the $z_{sp}$ image into account, the
emission lines are related to 321 possible counterparts in the
z$_{sp}$ image.  The average number of possible counterparts per
emission line is therefore two, which is consistent with one of these
being the real counterpart as randomly placed regions with the above
specified size cover on average one object in our $z_{sp}$ image.  For
each possible counterpart we have determined the associated line flux
from the spectroscopic image (which depends on the position of the
counterpart).

%For 101 objects the measured spectroscopic flux is larger than the
%imaging flux (which contains both line and possible continuum
%emission) ruling out that they are real counterparts.

Among the detected emission lines, 37 lines were part of an
identifiable line pair, either close (e.g.\
[\ion{S}{II}]$\lambda$6716,6731) or far apart (e.g.\ H$\beta$ at
4861\,\AA\ and [\ion{O}{III}]$\lambda$4959).  The corresponding 87
counterparts have been removed from the sample of possible LAEs.  As
the [\ion{O}{II}]$\lambda$3726,3729 doublet is only marginally
resolved in our slitless spectra, we have not eliminated candidate
[\ion{O}{II}] emitters.  From the remaining 234 objects, we have
selected candidate \lya\ emitters based on the singular and isolated
appearance of the line, and the absence of emission in the $I$ band or
strong $I - z_{sp}$ break of the possible counterpart.  There is only
one counterpart not detected in the $I$ band.  This object, KCS 1166,
has a line flux very close to the completeness limit (of
2.0$\times10^{-17}$\,\ecs) and an $I - z_{sp}$ colour $> 3.2$ or $>
1.0$ after subtraction of its line flux.  The corresponding emission
line has two other possible counterparts, which are almost equally
probable to cause the line (as they are close in position and
brightness) but do not show a spectral break between the $z_{sp}$ and
$I$ band (see Fig.\ \ref{direct} and Table \ref{counterparts}).  The
images of the three possible counterparts are rather faint for a
reliable determination of their sizes, but they are all consistent
with being unresolved.  The appearance of the emission line of this
LAE candidate is asymmetric but it is difficult to quantify this as
the s/n is low (see Fig.\ \ref{slitless_lya}).

%In the slit--mask employed, several other interesting objects were
%included, most notably objects detected on the $z_{sp}$ image not
%detected in the $I$ band and not associated with an emission line.
%These could be Lyman Break Galaxies
%\citep[LBGs,][]{1996ApJ...462L..17S} at $z \sim 6.5$ with weak or
%absent \lya\ emission or with a line at a wavelength just outside the
%narrow band filter.

% Among the $\sim 4500$ objects detected by SExtractor on the $z_{sp}$
% image, there are fourteen with $I - z_{sp} > 1.5$.  Four of these
% are possible counterparts of emission lines found in the slitless
% spectroscopy (including the two most promising candidates).  Objects
% with these colors are considered candidate Lyman Break Galaxies
% (LBGs, Steidel et al.\ 1996) at $z \sim 6$. Stanway et al.\ (2003)
% have carried out spectroscopy of 10 $(i' - z')_{AB} > 1.5$ objects,
% one of which is confirmed to be a galaxy at $z = 5.83$ while another
% is possibly at $z = 6.24$.

\begin{figure*}
   \centering
   \includegraphics[width=\linewidth]{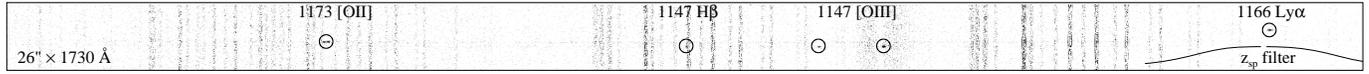}
   \caption{Complete two dimensional background subtracted spectrum of
   the slit with the sensitivity curve of the $z_{sp}$ filter indicated.}
   \label{complete_twodspec}%
   \end{figure*}

\begin{figure} \centering
   \includegraphics[width=\columnwidth]{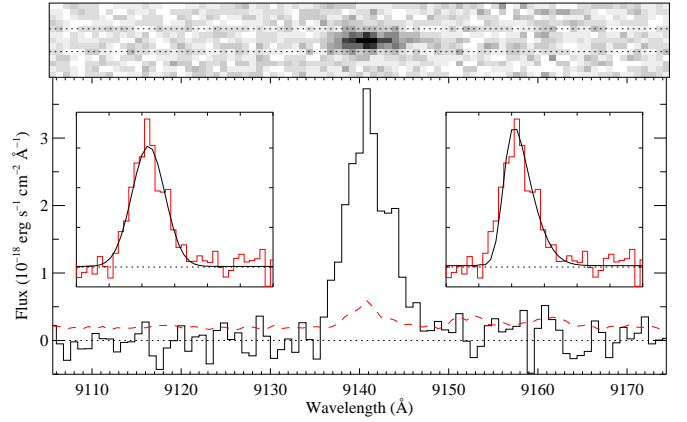} \caption{One and
   two dimensional slit spectrum of object KCS 1166.  The dotted lines
   indicate the borders of the extraction aperture (4 pixels or
   1\arcsec).  The dashed line indicates the signal to noise per
   pixel.  Shown as insets are two overlayed fits: a Gaussian and
   blue-absorbed Gaussian, both convolved with the Gaussian
   instrumental profile.}  \label{lya_fit} \end{figure}

\begin{figure}
   \centering
   \includegraphics[width=\columnwidth]{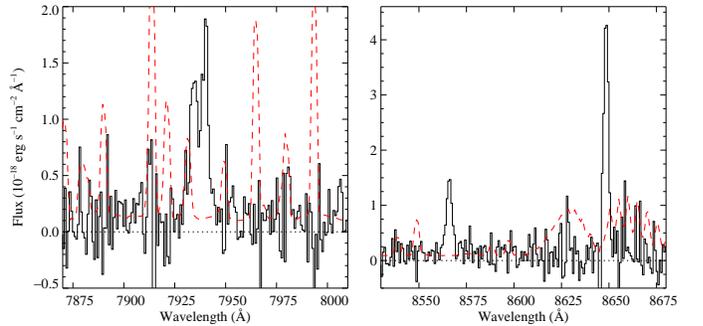}
      \caption{Slit spectrum of object KCS 1173 (left) and
      KCS 1147 (right).  The dashed line indicates the background
      divided by 50.}
      \label{slitmask_oiioiii}
   \end{figure}

To determine the actual counterpart of this emission line and to cover
a larger wavelength range with higher resolution and s/n, we have
carried out multi object slit spectroscopy.  Fig.\
\ref{complete_twodspec} shows the two dimensional spectrum obtained
with the slit mask, while Figs.\ \ref{lya_fit} and
\ref{slitmask_oiioiii} show the one dimensional spectra of the three
possible counterparts.  These spectra show that the line detected in
the slitless image is from KCS 1166.  The other two objects are also
line emitters: KCS 1173 is identified as an [\ion{O}{II}] emitter at
$z = 1.13$ and KCS 1147 as an [\ion{O}{III}] emitter at $z = 0.73$.  A
Gaussian fit to the emission line of KCS 1166 results in a central
wavelength of 9141.0\,\AA\ or $z = 6.518$ if identified with \lya.
KCS 1166 was also fit with a Gaussian profile truncated completely on
the blue side (to simulate absorption) whereafter it was (as all
fitted profiles) convolved with a Gaussian instrumental profile.  This
fit (also shown in Fig.\ \ref{lya_fit}) results in a FWHM of
280\,km\,s$^{-1}$ and redshift of 6.516 (for the unabsorbed line).
Parameters derived from the Gaussian profile fits are presented in
Table \ref{lines}.

Subtracting the line flux of KCS 1166 from the flux measured in the
$z_{sp}$ image, we obtain a line--free continuum of
1.6$\pm$1.1$\times10^{-20}$\,\ecs\,\AA$^{-1}$ or AB magnitude of 27.3.
In case the observed line can be identified with \lya, no continuum
flux is expected blueward of the line and the derived continuum
redward of the line would be
3.2$\pm$2.1$\times10^{-20}$\,\ecs\,\AA$^{-1}$.  The latter continuum
flux implies an observed EW of 600$\pm$400\,\AA\ for the line.  As the
continuum detection is only 1$\sigma$ and we have conservatively used
the high continuum flux, this value can be considered a lower limit.
We have also measured the continuum level in the spectrum of KCS 1166
below (9005 $< \lambda <$ 9030\,\AA) and above (9165 $< \lambda <$
9244\,\AA) the \lya\ line, resulting in values of -1.7$\pm$1.9 and
2.1$\pm$2.5$\times 10^{-20}$ \ecs\,\AA$^{-1}$, respectively.  The
difference of 3.8$\pm$3.1$\times10^{-20}$ \ecs\,\AA$^{-1}$ is
consistent with the continuum level redward of the line determined
from the imaging.

%Only one LBG candidate shows an emission line in the slit mask, but
%this line is accompanied by another, which are together identified as
%[\ion{O}{III}] lines at $z = 0.83$.

%                                     Two column figure (place early!)
%______________________________________________
%

%__________________________________________________ One column table
\begin{table}
  \caption[]{Possible counterparts of the emission line}
   \label{counterparts} 
\begin{tabular*}{\columnwidth}{r r r r r r}
\hline \hline
Source & \multicolumn{2}{c}{Coordinates (J2000)} & 
\multicolumn{1}{c}{$v$} & \multicolumn{1}{c}{$I$} & 
\multicolumn{1}{c}{\hspace{-1.3mm}$I - z_{sp}$} \\
\hline
KCS 1166 & 04 01 43.48 & -37 45 09.1 & $>$27.4 & $>$28.3 & $>$3.2 \\
KCS 1173 & 04 01 43.84 & -37 45 08.9 &    25.2 &    25.0 &    0.3 \\
KCS 1147 & 04 01 43.98 & -37 45 09.1 & $^{\mathrm{a}}$27.4 &    25.5 &    0.0 \\
\hline
\end{tabular*}
Magnitude limits are 1$\sigma$, errors are 0.1, except for
$^{\mathrm{a}}$ where it is 1.0
\end{table}

%
%                                                One column figure
%-----------------------------------------------------------
%
%______________________________________________________________

%
%                                                One column figure
%-----------------------------------------------------------
%
%______________________________________________________________

%__________________________________________________ One column table
\begin{table}
  \caption[]{Properties of the emission lines observed in the slit spectra}
   \label{lines} 
\begin{tabular*}{\columnwidth}{r r r r@{$\pm$}l r r}
\hline \hline
\multicolumn{1}{c}{Line} & \multicolumn{1}{c}{Source} & 
\multicolumn{1}{c}{$z$} & \multicolumn{2}{c}{Flux$^{\mathrm{a}}$} & 
\hspace{-1mm}FWHM$^{\mathrm{b}}$ & EW$_0$$^{\mathrm{c}}$ \\
\hline
\lya\,$\lambda$1216         & KCS 1166 & 6.518 & 1.9 & 0.1 &    160 & $>$80\\
\ion{O}{II}$\,\lambda$3726  & KCS 1173 & 1.129 & 0.5 & 0.1 &    110 &    10\\
\ion{O}{II}$\,\lambda$3729  & KCS 1173 & 1.129 & 0.6 & 0.1 &    110 &    15\\
H$\beta\,\lambda$4861       & KCS 1147 & 0.727 & 0.4 & 0.2 &     80 &    20\\
\ion{O}{III}$\,\lambda$4959 & KCS 1147 & 0.727 & 0.5 & 0.1 & $<$100 &    25\\
\ion{O}{III}$\,\lambda$5007 & KCS 1147 & 0.727 & 1.4 & 0.1 & $<$100 &    65\\
\hline
\end{tabular*}
$^{\mathrm{a}}$10$^{-17}$\,\ecs\hspace{0.2em}
$^{\mathrm{b}}$km\,s$^{-1}$ (deconvolved)
$^{\mathrm{c}}$\AA\hspace{0.2em}
\end{table}

\section{Line asymmetry}\label{sec:asymmetry}

Due to the absorption of \lya\ photons by intervening neutral
hydrogen, \lya\ lines are expected to have an asymmetric appearance in
constrast with non--resonant lines.  A reliable identification of an
emission line with \lya\ depends therefore critically on the
assessment of the line profile.  We have used various methods to
quantify the asymmetry of the line of KCS 1166.

First, we have applied the three tests presented in
\citet{2003AJ....125.1006R}. A \emph{model--free} test is provided by
symmetrizing the spectrum about an assumed line center and measuring
how well the symmetrized line fits to the observed line.  The $\chi^2$
value obtained by this method depends on the amount of continuum
included, which is relatively symmetric around the line and therefore
reduces the value of $\chi^2$.  It also depends on the error
estimation.  We have used the rms estimated from a source-free region
in the slitless image above and below the actual spectrum.  We have
measured the $\chi^2$ value for a range of assumed line centers in
steps of 0.1\,\AA.  The lowest reduced $\chi^2$ value measured was
1.35 for a conservative 20\,\AA\ wavelength range including some
continuum emission and 1.46 for a range of 10\,\AA\ which includes the
line only.  These values imply likelihoods of 1.4\% and 2.4\%,
respectively, that the line is consistent with its symmetrized
profile.  Two further tests attempt to measure the actual
asymmetry. For both, the wavelengths where the flux peaks and where it
has 10\% of its peak value on the blue and red side are determined.
The wavelengths are respectively: $\lambda_p$ = 9140.8\,\AA,
$\lambda_b$ = 9136.0\,\AA\ and $\lambda_r$ = 9147.2\,\AA.  The
\emph{wavelength ratio} is then defined as $a_\lambda = (\lambda_r -
\lambda_p) / (\lambda_p - \lambda_b)$ and the \emph{flux ratio} as
$a_f = \int_{\lambda_p}^{\lambda_r} f_\lambda d\lambda /
\int_{\lambda_b}^{\lambda_p} f_\lambda d\lambda$.  The resulting values
are $a_\lambda = 1.33 \pm 0.39$ and $a_f = 1.12 \pm 0.07$, where we
have assumed an error equal to the binsize (0.8\,\AA) for the
wavelength and the measured sigma for the flux.  These values are
consistent with those measured for other high redshift \lya\ lines,
according to \citet{2003AJ....125.1006R}.

\begin{figure} \centering
   \includegraphics[width=\columnwidth]{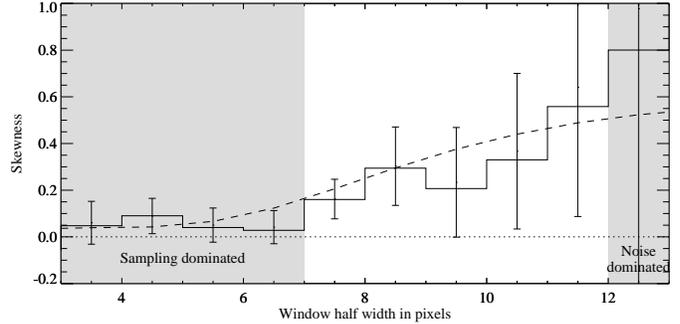} \caption{The
   skewness of the line for a range of window widths is represented as
   a histogram with one sigma errorbars. The dashed line represents
   the skewness of a Gaussian line profile, completely absorbed on the
   blue side and convolved with the instrumental profile.  Regions
   where the uncertainty is due to either sampling or noise
   are indicated.}\label{skewness} \end{figure}

In addition to these three tests, we have also measured the skewness
of the line for a range of wavelength windows.  The skewness or third
moment represents the degree of asymmetry of a distribution and is
positive for a line if the values above the central wavelength are
more spread out than those below it.  The result is presented in Fig.\
\ref{skewness}, where the uncertainties for each window are also
shown.  These uncertainties are a combination of noise and sampling.
For small window widths ($<$7 pixels), the uncertainties in wavelength
sampling dominate, while for large widths ($>$12 pixels), the
uncertainties in flux dominate.  In between (7 -- 12 pixels), a
positive skewness is measured, consistent with a Gaussian emission
line completely absorbed on the blue side and convolved with the
instrumental profile.

%The results of these tests show what is already clear from a visual
%inspection of the spectrum: there is an indication that the line is
%asymmetric but the signal--to--noise is low.  Part of the asymmetry is
%contained in the \emph{body} of the line, as shown by $a_f$ and part
%of it is contained in the red tail, as shown by the skewness.

The results of all tests are consistent with the line being
asymmetric, strongly suggesting that the line has excess emission on
its red side as expected for a high redshift \lya\ line.

%
%                                                One column figure
%-----------------------------------------------------------
%
%______________________________________________________________

\section{Discussion}\label{sec:discussion}

The absence of flux below the emission line at 9141\,\AA, especially
in the deep $I$ band image (see Table \ref{counterparts}) and the
observed asymmetry support the identification of this line with \lya\
at $z = 6.518$.  The flux ratio between the line--free $z_{sp}$ and
$I$ bands of $\sim$2 (or $\sim$4 for continuum redward of the line
only) is consistent with strong attenuation by the Lyman forest at
high redshift, but difficult to reconcile with a 4000\,\AA\ or Balmer
break.  The identification with other lines than \lya\ is very
unlikely.  As the slit spectrum covers wavelengths up to 9262\,\AA, we
can exclude the identification with [\ion{O}{III}] based on the
absence of other lines in the spectrum.  For the same reason, the
identification with H$\alpha$ is improbable as the accompanying
[\ion{N}{II}] lines would have to be ten times fainter and the
H$\alpha$ EW$_0$ would be in excess of 400\,\AA.  We would have
resolved line doublets as [\ion{O}{II}], \ion{Mg}{ii} and \ion{C}{iv}
(as shown by the spectrum of KCS 1173, Fig.\ \ref{slitmask_oiioiii})
and, if identified with [\ion{O}{II}], the line would have an EW$_0 >$
240\,\AA, which is a factor of two higher than the largest EW$_0$
observed in low redshift samples of emission line galaxies
\citep[e.g.\ ][]{1996A&AS..120..323G}.  In addition, the measured $I -
z_{sp}$ colour, after line subtraction, is more consistent with a
Lyman break than a Balmer jump.

We can derive the SFR for the \lya\ emitter based on the conversions
for \ha\ and UV continuum given in \citet{1998ARA&A..36..189K} and
assuming a \lya/\ha\ ratio of 8.7 \citep{1971MNRAS.153..471B}.  The
\lya\ luminosity of the line is 1.1$\times10^{43}$ erg s$^{-1}$, which
results in a SFR$_{\rm Ly\alpha}$ of 10.0$\pm$0.5
M$_\odot$\,yr$^{-1}$, while the continuum luminosity density of
6.6$\times10^{-28}$\,erg s$^{-1}$\,Hz$^{-1}$ results in a SFR$_{\rm
UV} > 9\pm$6 M$_\odot$\,yr$^{-1}$.  Note that the SFR$_{\rm UV}$
conversion used is calibrated only for the range between 1500 and
2800\,\AA, while our UV continuum is measured near 1200\,\AA, which
can add additional uncertainties.

The useful surface area of the survey is 6\farcm18$\times$6\farcm93 or
42.8 square arcminutes.  We do not take into account the surface area
covered by spectra of bright objects which may inhibit detection of
underlying emission lines as this is only a very small fraction (about
1\%).  The surface density of the complete sample of LAEs (with fluxes
$\gtrsim 2\times10^{-17}$\,\ecs) in our field is 0.023$^{+0.054}_{-0.019}$
arcmin$^{-2}$.  Given the comoving volume of 18460 Mpc$^3$ for our
survey, we derive a lower limit to the SFRD at $z = 6.5$ of
5$\times10^{-4}$ M$_\odot$ yr$^{-1}$ Mpc$^{-3}$ (derived from the
\lya\ flux).  Using a much larger field of view and including LAEs
with fluxes $> 0.90\times10^{-17}$\,\ecs, \citet{2003PASJ...55L..17K}
arrive at a comparable SFRD of 4.8$\times10^{-4}\, h_{0.65}$
M$_\odot$\,yr$^{-1}$\,Mpc$^{-3}$.  This SFRD is about ten times higher
than that estimated for the complete sample of LAEs at $z = 6.5$ with
fluxes $> 2\times10^{-17}$\,\ecs\ in the LALA survey
\citep{Rhoads2004}.  It is interesting to note that the SFRD derived
from our observations is comparable with that derived from the
abundance of $i'$-band drop-outs in the Hubble Ultra Deep Field
\citep{Bunker2004}, which low value poses a challenge for models
suggesting that the bulk of star forming galaxies that reionized the
universe lie at redshifts just beyond $z = 6$.  Larger samples of LAEs
at $z = 6.5$ are needed to measure the clustering of these galaxies
and the resulting cosmic variance to obtain a better view of the SFRD
at $z > 6$.  

%\citep{1986ApJ...303..336G}% Gehrels et al.
%6'11'' * 6'56'' = 6.18x6.93 arcsec, 6.443 < 6.608

One of the major challenges in observational cosmology is the
detection of the first stars.  These enigmatic objects may be
responsible for the ionization of neutral hydrogen in the universe and
should be formed from primordial matter.  Although recent WMAP results
seem to place reionization at $z > 10$ \citep{2003ApJS..148..161K},
the existence of metal-free stars at lower redshift depends strongly
on the distribution of supernova products.  The detection of large
optical depth neutral hydrogen in front of $z \sim 6$ QSOs
\citep[e.g.\ ][]{2002AJ....123.2151P} illustrates that the ionization
process might have occured very inhomogenously throughout the
universe.  \citet{2003ApJ...589...35S} predict that metal-free stars
can be formed down to $z < 4$ and might be found in presently known
LAEs \citep[see also][]{2003ApJ...596..797F}.  The \lya\ EW of PopIII
objects is predicted to be very high \citep[400 -- 800\,\AA, ][]
{2003A&A...397..527S}, although the observed EW can be diminished by
\ion{H}{I} absorption and depends therefore on the geometry of the
ionized and neutral gas.  The lower limit to the EW$_0$ of KCS 1166
(80\,\AA) is not particularly high but we stress that the unabsorbed
EW$_0$ can be significantly higher.  Table \ref{z6.5LAEs} shows the
EW$_0$s for all known $z = 6.5$ LAEs, as computed assuming that the
observed continuum flux (or its 1$\sigma$ upper limit) is only present
redward of the emission line.  The galaxy described by
\citet{Rhoads2004} and the galaxy described in this work have the
highest EW$_0$s and are therefore the most likely LAEs at $z = 6.5$ to
contain metal-free stars.  Given the present observing capabilities
and the uncertain \lya/\ion{He}{ii} flux ratio for PopIII objects
\citep{2003A&A...397..527S}, it would be just feasible to detect a
strong \ion{He}{ii} line, a unique feature of metal-free stellar
populations.  Further investigation is needed to determine the nature
of the stellar population of this and other $z = 6.5$ galaxies.

%__________________________________________________ One column table
\begin{table}
  \caption[]{Properties of known \lya\ emitters at $z = 6.5$}
   \label{z6.5LAEs} 
\begin{tabular*}{\columnwidth}{l r r r r r}
\hline \hline
\multicolumn{1}{c}{Name} & 
\multicolumn{1}{c}{$z$} & \multicolumn{1}{c}{Flux$^{\mathrm{a}}$} & 
EW$_0^{\mathrm{b}}$ & \multicolumn{1}{c}{Ref$^{\mathrm{d}}$} \\
\hline
HCM 6A                    & 6.556 & 0.65 & 25      & (1)\\
SDF\,J132415.7+273058     & 6.541 & 2.06 & 21      & (2)\\
SDF\,J132418.3+271455     & 6.578 & 1.13 & 44      & (2)\\
LALA\,J142442.24+353400.2 & 6.535 & 2.26 & $>$85$^{\mathrm{c}}$ & (3)\\
KCS 1166                  & 6.518 & 1.90 & $>$80$^{\mathrm{c}}$ & (4)\\
\hline
\end{tabular*}
$^{\mathrm{a}}$10$^{-17}$\,\ecs\hspace{0.2em}
$^{\mathrm{b}}$\AA\hspace{0.2em}
$^{\mathrm{c}}$1$\sigma$ lower limit
$^{\mathrm{d}}$ (1) \citet{2002ApJ...568L..75H}, (2)
\citet{2003PASJ...55L..17K}, (3) \citet{Rhoads2004}, (4) This work
\end{table}

%\begin{acknowledgements}
%\end{acknowledgements}

\bibliographystyle{aa}
\bibliography{Ge171}

\end{document}